# Design and study of a coronavirus-shaped metamaterial sensor stimulated by electromagnetic waves for rapid diagnosis of covid-19

Yadgar I. Abdulkarim[1,2], Halgurd N. Awl[3], Fahmi F. Muhammadsharif[4], Karzan R. Sidiq[5], Salah Raza Saeed[5], Muharrem Karaaslan[6], Shengxiang Huang[1], Heng Luo[1], Lianwen Deng[1],*

[1]School of Physics and Electronics, Central South University, Changsha, Hunan 410083, China.
[2]DepartmenPhysics Department, College of Science, University of Sulaimani, Sulaimani, 46001, Iraq.
[3]Department of Communication Engineering, Sulimani polytechnic University, Sulaimani, 46001, Iraq.
[4]Department of Physics, Faculty of Science and Health, Koya University, 44023 Koya, Iraq
[5]Charmo Centre for Research, Training and Consultancy, Charmo University, 46023 Chamchamal, Iraq
[6]Department of Electrical and Electronics, Iskenderun Technical University, Hatay, 31100, Turkey

Corresponding author: (e-mail: denglw@csu.edu.cn).

**ABSTRACT** We propose a new technique of utilizing metamaterials-based sensor for rapid diagnosis of covid-19 through electromagnetic-stimulated analysis of the blood drawn from the patient. The sensor was inspired by a coronavirus in plane-shaped design with presume that its circular structure might produce a broader interaction of the electromagnetic waves with the blood sample. The sensor was designed numerically and tested experimentally by evaluating variations in the reflection coefficient (S11) and transmission coefficient (S21) of the waves at resonant frequency. Results of covid-19 relevant blood sample showed a pronounced shift in the main resonant frequency of about 740 MHz compared to that of the control blood sample. We believe that with the help of the proposed sensor a significant breakthrough can be achieved for rapid diagnosis of covid-19 within few seconds.

## I. INTRODUCTION

The pandemic coronavirus diseas-2019, commonly called COVID-19, is caused by a novel strain of coronavirus that recognized as Sever Acute Respiratory Syndrome Coronavirus2 (SRAS-CoV-2) [1, 2]. The virus is a member of Coronaviridae family, which contains enveloped single stranded RNA viruses. The SRAS-CoV-2 primarily attacks the respiratory system of human and occasionally causes severe pneumonia. The sever cases is also followed by multi-organ dysfunctions, respiratory failure, and death[3]. The COVID-19 was first reported in Wuhan-Hubei province of China in December 2019, soon after the disease occurred in different countries around the world [4, 5]. So, COVID-19 was announced as a pandemic disease by World Health Organization (WHO) in March 2020[6]. Up to 23rdJune 2020, about9 million cases and almost half a million deaths were reported globally [7]. Despite tight medical actions by the countries, these data reveal that further preventive, diagnostic and therapeutic measurements are still required so as to control the pandemic as soon as possible. Currently, there are both immunological and molecular laboratory tests for the detection of COVID-19. However, the immunological antigen-antibody test may have sensitivity and specificity drawbacks [8], while the molecular polymerase chain reaction (PCR) has the disadvantages of lacking the PCR infrastructure in every hospitals, cost and time-consuming [8, 9]. The quick ascending morbidity of COVID-19 makes alternative diagnostic techniques necessary. Consequently, the future developed diagnostic methods should be more sensitive, cheaper and faster in order to cope with the increasing number of COVID-19 cases around the world.

It is well known that the human blood is composed of plasma and cells. Plasma is the liquid portion of blood and a mixture of water, proteins (albumin, clotting factors, antibodies, enzymes, and hormones) and other dissolved materials. The blood cells are erythrocyte (red blood cell), leukocytes (white blood cells) and thrombocytes. The leucocytes include neutrophils, basophils, eosinophils, monocytes and lymphocytes. These white cells are related to body defense against pathogens [10]. Studies observed alterations in the biochemistry and cellular components of blood in COVID-19 patients. The biochemical components of blood such as Alanine aminotransferase (ALT), Aspartate aminotransferase (AST) C-reactive protein (CRP), lactate dehydrogenase (LDH), and Urea significantly increased, while albumin decreased in the serum of the PCR positive COVID-19 patients [11-14]. Unlike most bacterial and viral infections that generally cause high leukocyte [15], and lymphocyte counts [16], significant decreases occurred in the number and percentage of leukocytes and lymphocytes in PCR confirmed cases of COVID-19 [11, 12, 14, 17-20].Therefore, developing a technique to detect the changes in these unique blood parameters could be a significant achievement for rapid diagnosis of COVID-19.

A review of literature showed that a pronounced variation in the dielectric permittivity of blood is occurred when counts of leukocytes, lymphocytes, as well as IgG and IgM antibodies in are altered [21-23]. Hence, based on the aforementioned evidences that leukocytes and lymphocytes in the blood of Covid-19 patients are dramatically reduced, we have established the hypothesis that it is possible to detect changes in the dielectric permittivity of blood among covid-19 patients by the use of electromagnetic inspired metamaterial sensors. The involvement of metamaterials (MTMs) in the construction of highly sensitive sensors has become a new trend. This is due to the periodical structure of the MTMs, which acts to manipulate the electromagnetic waves behavior and hence producing unique features such as negative refractive index, permeability and permittivity [24, 25].

We have previously reported various designs and shapes of metamaterial-based sensors for the detection of different liquids and chemical materials [24-26]. The working principle of these sensors is based on the shift of resonant frequency in the transmitted electromagnetic signals through the samples due to their difference in permittivity and permeability values. In this study and for the first time, we propose MTM-based sensor for rapid detection and diagnosis of covid-19. The selected sensor design was inspired by a coronavirus-shaped architecture with a presume that this kind of circular structure may lead to a broader interaction of the electromagnetic waves with the blood sample. The proposed sensor was designed and tested both numerically and experimentally by evaluating variations in reflection coefficient ($S_{11}$) and transmission coefficient ($S_{21}$) at resonant frequency. Results showed that the proposed sensor can be a viable tool for rapid detection and diagnosis of covid-19 patients within few seconds.

## II. DESIGN AND FABRICATION OF THE PROPOSED SENSOR

The corona virus-shaped microwave sensor was designed and numerically investigated by CST microwave suite based on the Finite Integration Technique (FIT). The sensor incorporates a corona-virus shaped resonator connected with a transmission line, as shown in Figure 1, where P1 is the first port, P2 is the second port, L is inductance of the CRR, R is resistance of the CRR, C1 is capacitance between the outer CRR and inner conductor and C2 is capacitance between the inner conductor and ground plane. The layout of the proposed sensor, composed of three layers, along with its equivalent electrical circuit, is shown in Figure 2. On the top layer, a closed ring resonator, with star-shaped conductor inside it, is joined with two transmission lines. The top layer is made of copper with a conductivity of $5.8 \times 10^7$ S/m and thickness of 0.035 mm, while both ends of the transmission lines are excited by discrete port 1 and 2. The second layer is FR-4 substrate with a thickness of 1.6 mm, dielectric constants and loss tangent of 4.3 and 0.02, respectively. This substrate layer has been chosen due to the low loss value and availability in our labs. The bottom layer is fully covered with a copper metal of 0.035 mm thick, which behaves as a ground plate for the structure. The space between closed circular ring (CCR) and the conductive inner star shape is utilized as the sensor layer to be filled with blood samples under investigation.

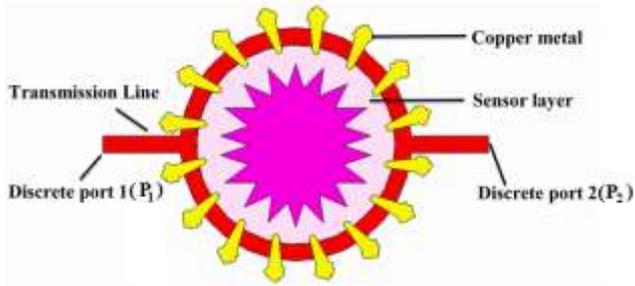

**FIGURE 1**: Front view of the proposed metamaterials-based coronavirus-shaped resonator.

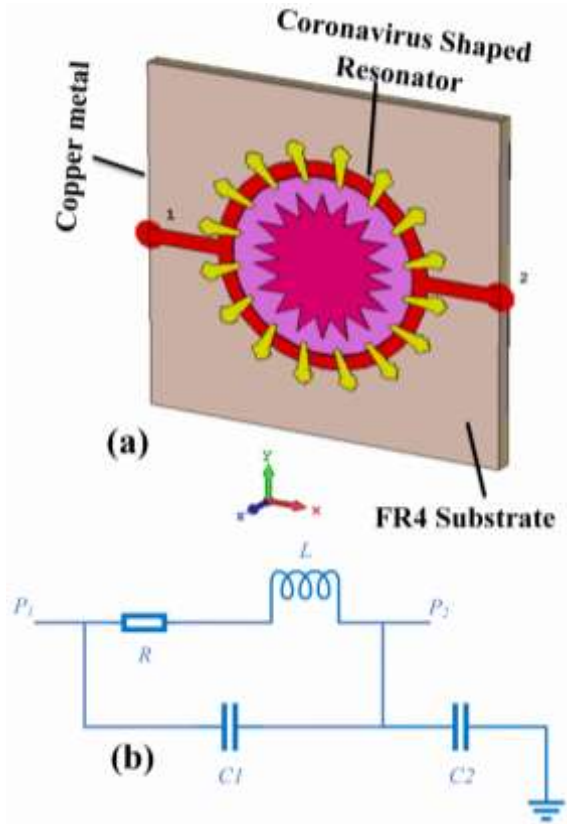

**FIGURE 2:** (a) Prospective view of the proposed metamaterials-based sensor and (b) its equivalent circuit diagram.

The coronavirus-shaped resonator was fabricated by using LPKF ProtoMAT E33 prototyping machine with the same dimensions of the numerically designed one and tested in the frequency range from 4 to 8 GHz, as shown in Figure 3. Two coaxial test cables were connected to the metamaterial sensor via two ports, as shown in Figure 3-b. Before taking measurements, the vector network analyzer (VNA) was connected to the structure through the coaxial cables and then calibration has been done in three steps of open circuit, short circuit and 50 Ohm load connector in the desired frequency range, as shown in Figure 3-c. To take the reference data, simulation and experiment were carried out with the presence of air and three blood samples in the sensor layer.

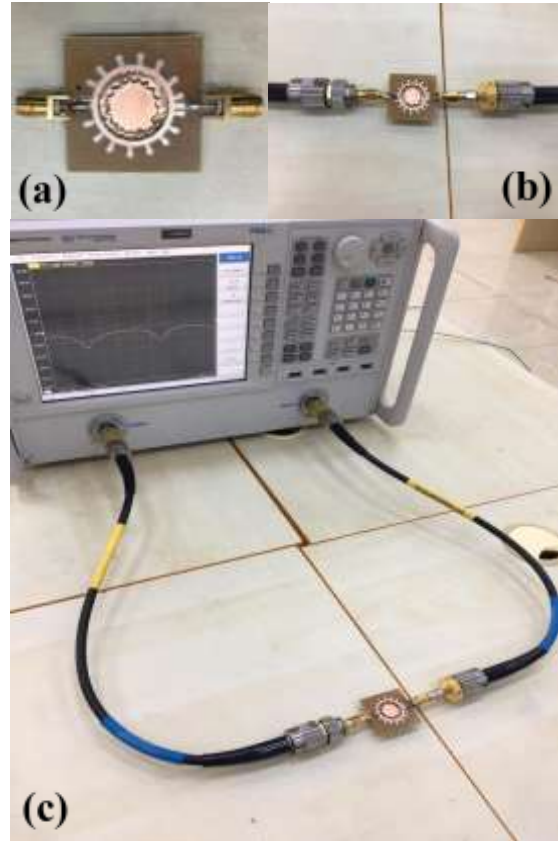

**FIGURE 3**: (a) Photograph of the proposed structure (b), its two-port connection and (c) experimental set up.

### IV. RESULTS AND DISCUSSION

The reflection (S11) and the transmission (S21) coefficients in magnitude and in decibel were monitored to investigate the interaction of electromagnetic (EM) field with the designed sensor over the frequency range of 4-8 GHz (C-band), as shown in Figure 4. It was seen from the results that there exist three minima and maxima peaks at the resonant frequencies of 4.75 GHz, 6.5 GHz, and 7.75 GHz for reflection and transmission coefficient parameter of designed sensor, respectively with magnitude of -14 dB, -13.5dB and -13.3 dB, respectively. This makes the sensor to be efficiently used for COVID-19 diagnoses over this wide range of frequency (C-band).

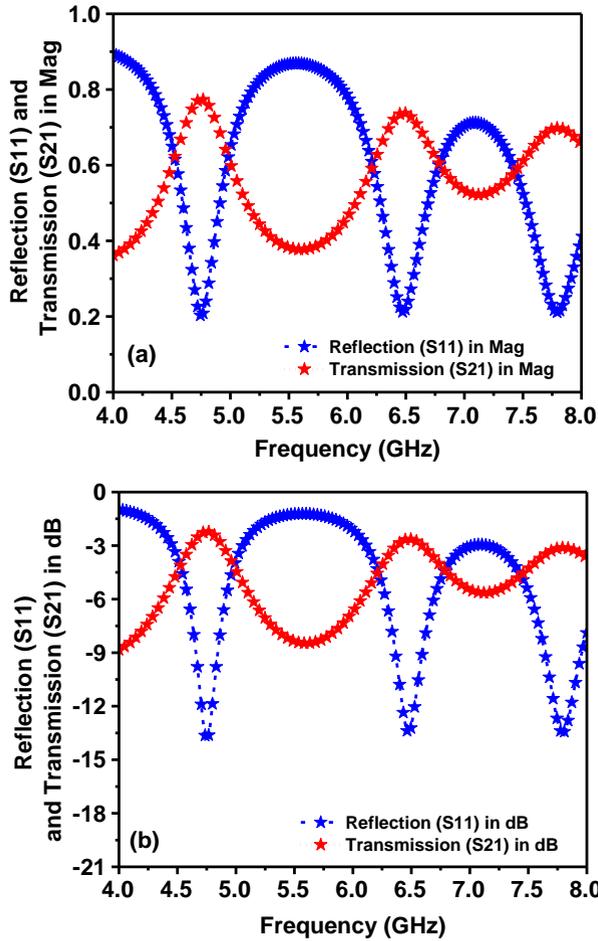

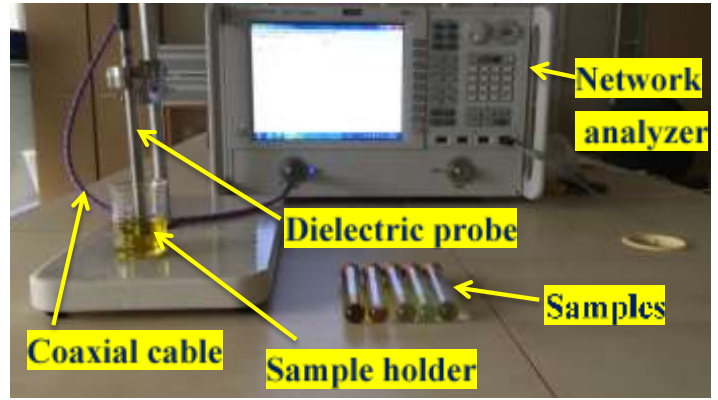

FIGURE 5: Experimental setup to perform dielectric measurement of the normal blood samples using 85070E dielectric probe kit.

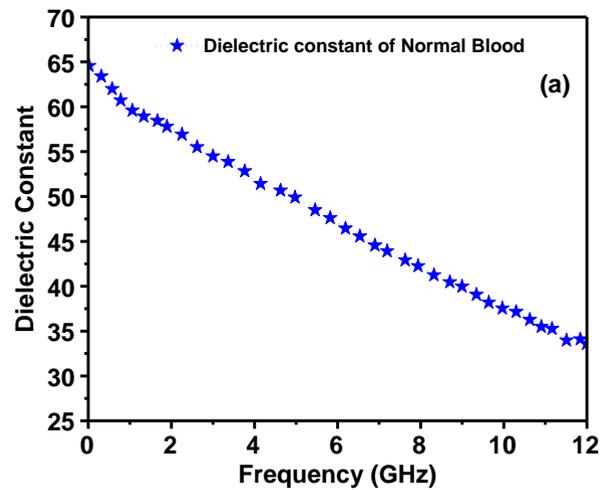

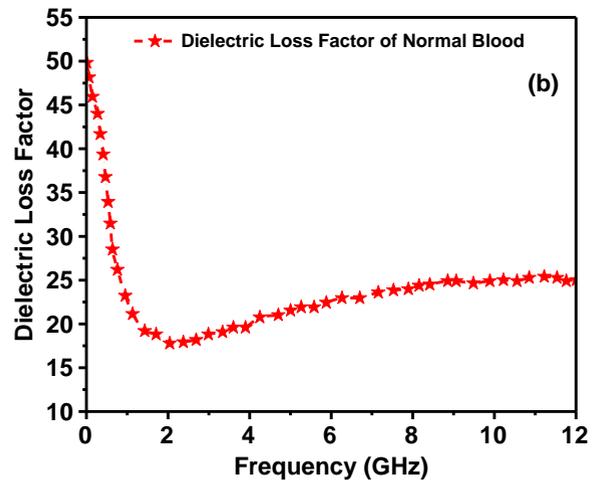

FIGURE 4: (a) Simulated results of reflection and transmission coefficient for the proposed coronavirus-shaped sensor: (a) in magnitude and (b) in dB.

FIGURE 6: Measured dielectric constant (a) and dielectric loss factor (b) of normal blood.

The measured dielectric permittivity and loss tangent for air and normal blood were recorded and data were imported into the CST software to estimate the reflection coefficient and transmission coefficient. To measure the dielectric constant and dielectric loss factor of normal blood, a dielectric probe kit device was utilized in conjunction with vector network analyser (VNA) Agilent 85070E, as shown in Figure 5. The measured electrical properties of normal blood in the frequency range 0-12 GHz are shown in Figure 6. Initially, the dielectric probe was calibrated by using air and pure water of known electromagnetic parameters [27]. Table 1 shows the parameters values at specified frequencies. One can see from the results that the dielectric constant has decreased linearly with the increase of frequency. However, the dielectric loss factor was exponentially decreased from 0 to 2 GHz and then increased gradually to reach a plateau shape at high frequencies from 9 GHz onwards.

TABLE 1
THE MEASURED DIELECTRIC PARAMETERS OF NORMAL BLOOD AT
SELECTED FREQUENCIES.

| Frequency (GHz) | dielectric constant | Dielectric Loss factor |
|---|---|---|
| 0 | 64.9 | 50 |
| 1 | 60 | 21.2 |
| 2 | 57.5 | 17 |
| 3 | 54.8 | 18 |
| 4 | 52 | 19.9 |
| 5 | 50 | 22 |
| 6 | 47.5 | 22.5 |
| 7 | 44.9 | 23 |
| 8 | 42.5 | 24 |
| 9 | 40 | 25 |
| 10 | 37.6 | 25 |
| 11 | 35.1 | 25.1 |
| 12 | 34 | 25 |

Regarding the blood of COVID-19 patient (COVID-19 blood henceforth), data of dielectric constant and loss tangent was calculated based on their decrement by 5% compared to that of the normal blood at high frequency, i.e. the plateau region. It was evidenced from literature that dielectric permittivity of blood is decreased when the count of lymphocytes is increased [21, 22]. Ermolina et. al. used different mixture formulas to conclude that the relationship between static dielectric permittivity of blood as functions of volume fraction of cells is almost linear from 0% to about 20% of cells volume fraction [22]. It was reported that in most cases of COVID-19, lymphocyte percentage (LYM%) was reduced to lower than 5% [28]. Furthermore, when the immunoglobulin/antibody proteins are just started to produce by the immune system in response to the COVID-19 antigen, the dielectric permittivity is further lowered. This is because it was evidenced that the increase of IgG and IgM immunoglobulins has led to the decreased capacitance and dielectric permittivity of blood [23]. Hence, 5% smaller values of dielectric permittivity and loss tangent compared to those of the normal blood were imported into the CST software to represent the electrical response of COVID-19 blood. Figure 7 shows the reflection coefficient for air, normal blood and COVID-19 blood samples. It can be seen that the resonant frequency has three different peaks at 4.38 GHz, 6.26 GHz and 7.82 GHz corresponding to the reflection -13 dB, -15.2 and -29 dB, respectively. Interestingly, for the covid-19 samples, the resonant frequency presented two peaks at 5.1GHz and 7 GHz, with a pronounced shift of about 720 MHz and 740 MHz, respectively. We believe that with the help of the proposed sensor a significant breakthrough can be achieved for rapid diagnosis of covid-19 within few seconds.

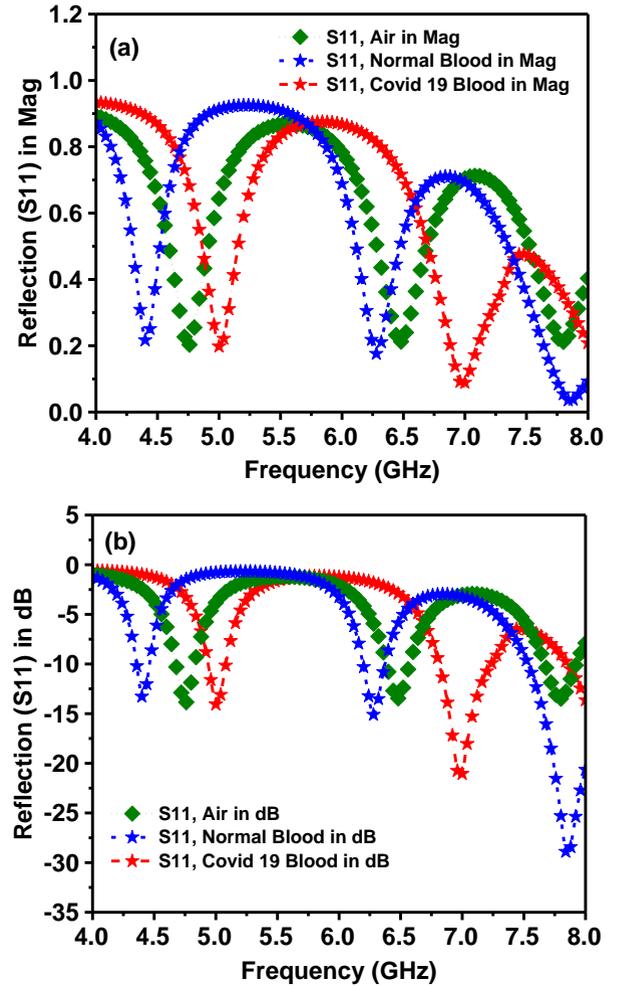

**FIGURE 7**: Simulated results for air, normal blood and Covid-19 blood samples: (a) reflection ($S_{11}$) in magnitude and (b) reflection ($S_{11}$) in decibel.

Similarly, the transmission coefficient of the sensor layer filled with COVID-19 blood showed a significant shift in the resonant frequency compared to that of air and normal blood-filled sensor, as shown in Figure 8. In the case of COVID-19 blood, the frequency maxima have shifted towards higher frequency by about 620 MHz and 750 MHz, respectively. This suggests that the proposed structure can be highly sensitive for the detection of COVID-19.

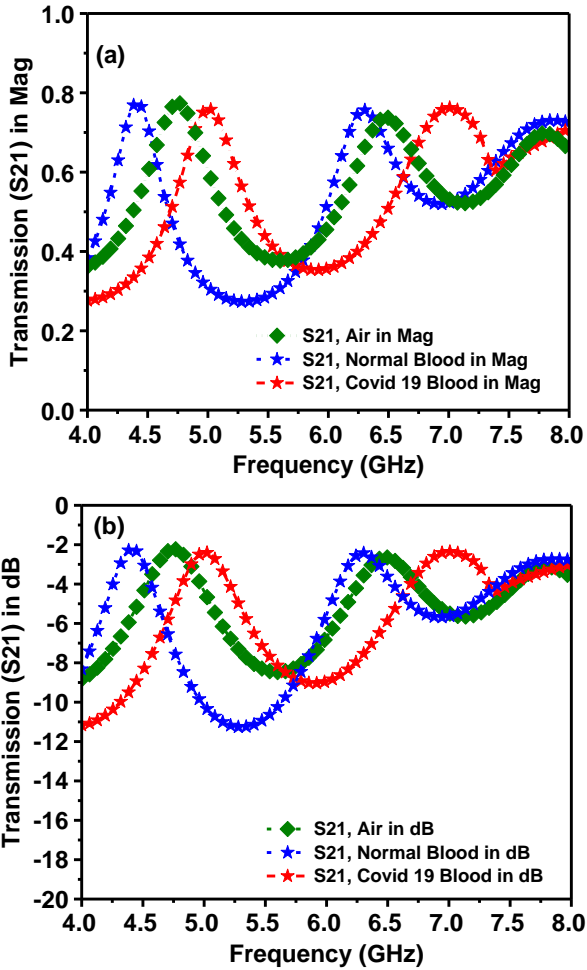

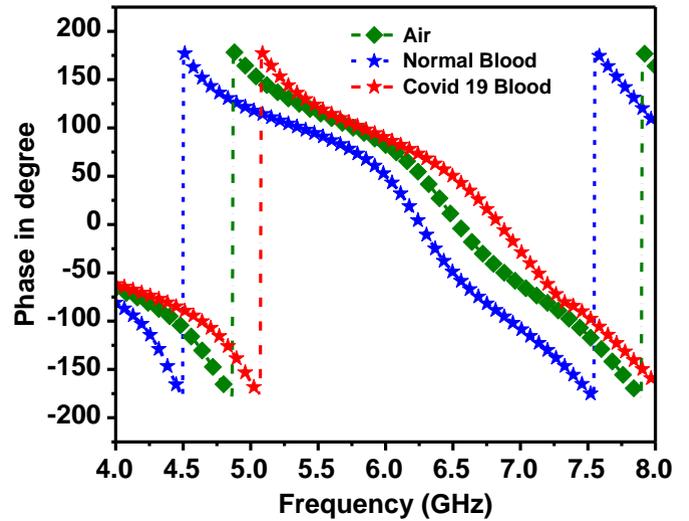

**FIGURE 9:** Simulated transmission phase of the sensor tested for samples of air, normal blood and COVID-19 blood.

**FIGURE 8**: (a) Transmission coefficient ($S_{21}$) spectra of air, normal blood and COVID-19 blood samples in magnitude (a) and in decibel (b).

Figure 9 shows the simulated result of phase variation in the transmitted wave recorded in the frequency range from 4 to 8 GHz when different samples were inserted into the sensor layer. One can notice from the figure that the degree of phase variation is highly deviated for each of the samples under test. Consequently, the blood of patients with covid-19 has resulted in a pronounced phase difference compared to that of the normal blood, implying that the sensor can be highly efficient for the detection of covid-19 cases.

In order to perform parametric study on the proposed sensor, thereby optimizing its geometrical dimensions, the built-in genetic algorithm in the CST software was used with empty sample. The effect of sensor layer radius (Rs), width of the transmission line (WT) and radius of the outer ring (Ro) on the resonant frequency in both reflection (S11) and transmission (S21) mode was monitored, as shown in Figure 10. As such, a clear shift in the resonant frequency was observed when the radius increased from 8.4 mm to 9.6 mm. The important part of the designed structure is the sensor layer because when the samples are put inside it, the resonant frequency can be shifted with respect to the change in electrical properties of the samples. Even a small variation in the electrical properties of the samples has resulted in a clear shift in the resonant frequency, which might be because of the symmetrical structure of the sensor. Results showed that the optimum value of radius of the sensor layer is 9 mm.

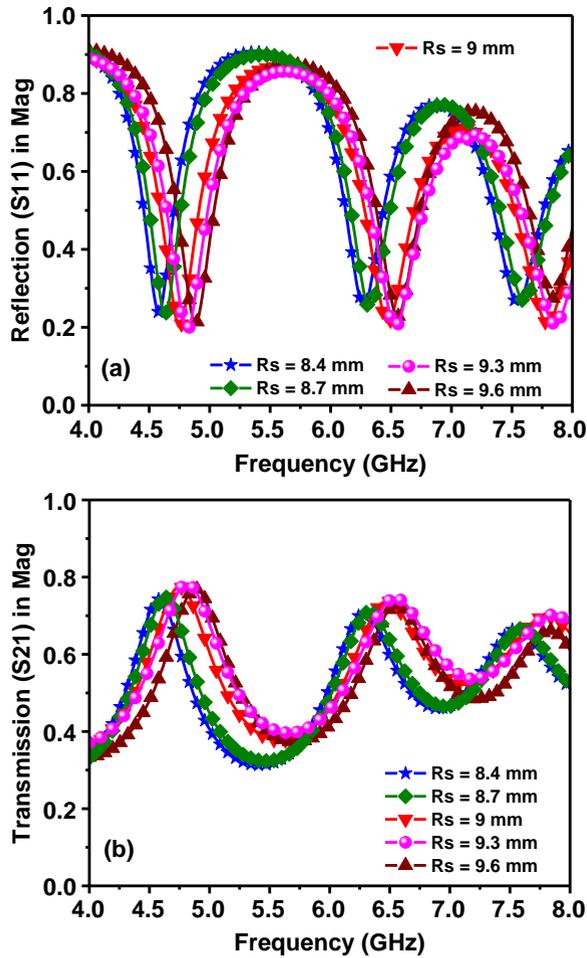

**FIGURE 10**: Effect of sensor layer radius on the resonant frequency of (a) reflection and (b) transmission spectra.

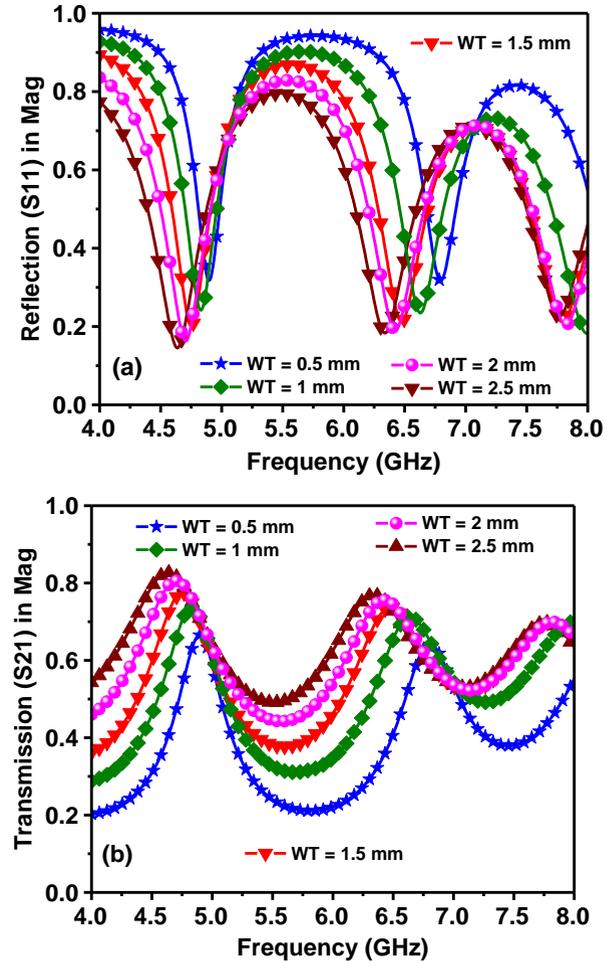

**FIGURE 11**: Effect of transmission line width on the resonant frequency of (a) reflection and (b) transmission spectra.

Figure 11 shows the effect of varying width of the transmission line on the resonant frequency of the reflection and transmission coefficient spectra of the sensor. Noticeably, the variation width of the transmission line has affected on the resonant frequency, where the optimum vale of the transmission line width was set to be 1.5 mm. Results indicated that the thinner the transmission line, the lower transmission coefficient and vice versa. It was seen from the Figure 11(b) that there exist three maxima peaks at the resonant frequencies of 4.75 GHz, 6.5 GHz, and 7.75 GHz for the transmission coefficient with the bandwidth is about 1.75 GHz.

.

The final parametric study was taken by changing the radius of the outer ring resonator from 9.5 mm to 11.5 mm in steps of 0.5 mm, while other dimensions were kept constant. The optimum value of this parameter was found to be 10.5 mm, at which the reflection coefficient presented three resonant peaks at 4.75 GHz, 6.5 GHz and 7.80 GHz, as shown in Figure 12. It is worth to mention that at a small radius of 9.5 mm, resonant frequency resulted in a significant shift of about 4.25 GHz, 5.75 GHz and 7 GHz for the first, second and third resonance peaks, respectively. Consequently, the total shift in the resonant frequency is about 500 MHz, 750 MHz and 800 MHz. When the outer radius of the resonator was set to 9.5 mm and 10 mm, there was a clear shift seen. This can be attributed to the overall dimension of the resonator and that the waveguide wall is higher than the other dimensions of the coronavirus resonator

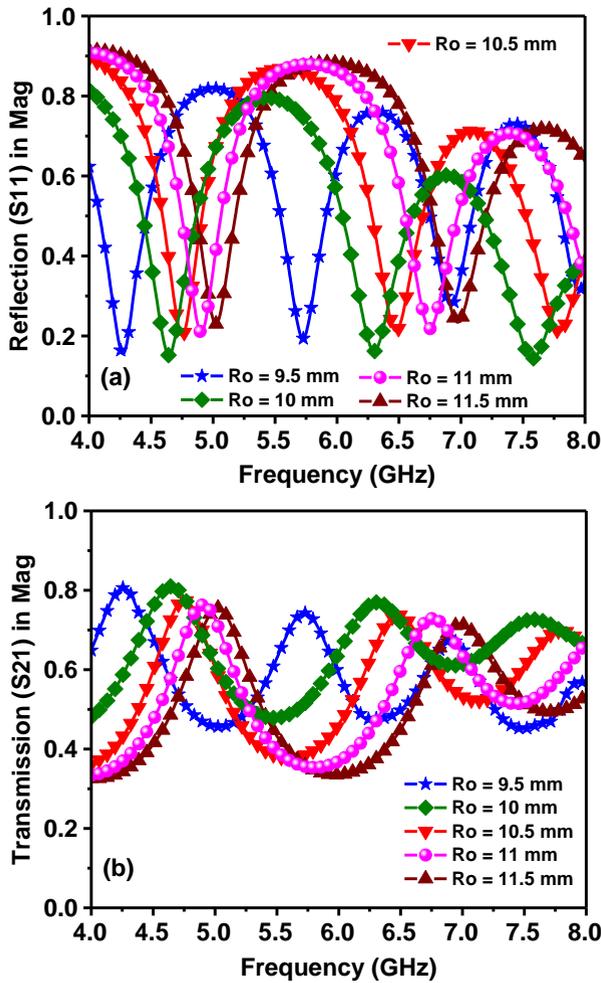

**FIGURE 12:** Effect of the width of outer ring resonator radius on the resonant frequency of (a) reflection and (b) transmission spectra.

The electrical properties, lactate, glucose level and temperature of the three different blood samples used in this work are shown in table 2. Figure 13 shows the measured reflection and transmission coefficient of the sensor with and without the presence of normal blood samples. Interestingly, the performance of the sensor in the low frequency range and at the main resonant frequency of 4.75 GHz is highly stable in terms of results reproductively. It was seen that the resonant frequency position remained unchanged when three samples of normal blood were tested. Hence, we believe that the proposed sensor can be efficiently utilized for rapid detection of covid-19 within few seconds. In this way, the use of metamaterial sensors can be a significant breakthrough for rapid covid-19 diagnosis.

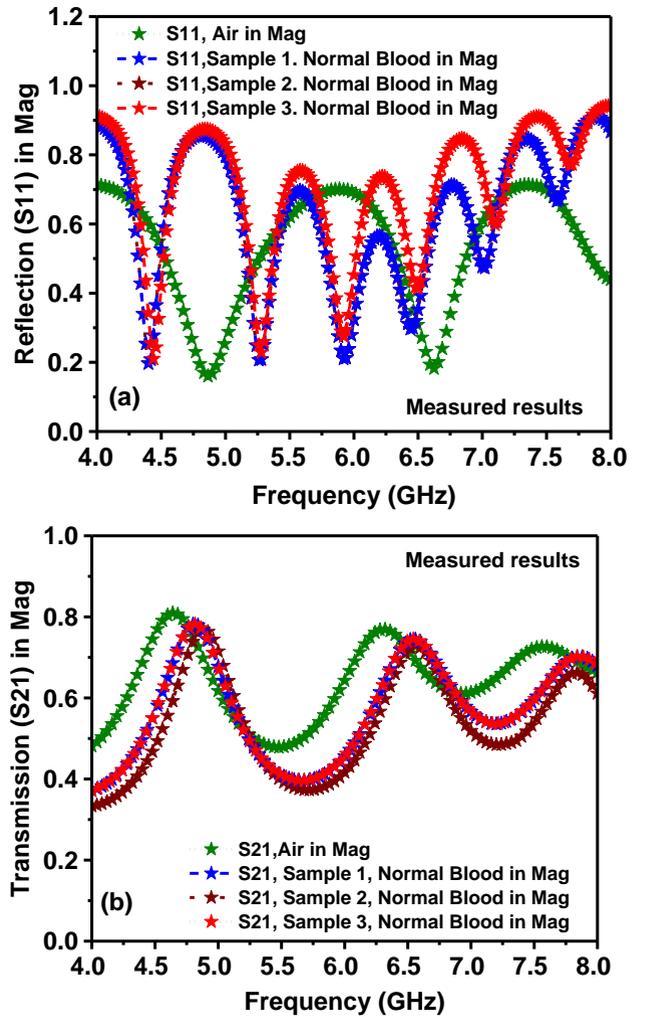

**FIGURE 13**: The measured reflection coefficient (a) and transmission coefficient (b) spectrum for empty sensor and filled sensor with normal blood.

Figure 14 shows the surface current distribution across the proposed sensor at resonance frequency of 4.75 GHz. It was observed that the current flow is bidirectional from the centre of the resonator to the ports and vice versa. The current distributing on the outer ring resonators emerges from the outside and enters at the resonator's junction. Consequently, the parallel and antiparallel current flowing is responsible for controlling the electric and magnetic responses, respectively. In conclusion, the electric and magnetic response of the whole sensor can be interpreted to detect various samples based on their intrinsic dielectric parameters.

TABLE 2

PROPERTIES OF THE BLOOD SAMPLES USED IN THE STUDY.

| Properties | Blood #1 | Blood #2 | Blood #3 | STDEV |
|---|---|---|---|---|
| Dielectric constant | 64.9 | 63.7 | 64.5 | ±0.61 |
| Dielectric loss factor | 50 | 49.6 | 49.8 | ±0.20 |
| Temperature (0C) | 29.6 | 29.8 | 29.7 | ±0.10 |
| Glucose (mmol/l) | 4.46 | 4.45 | 4.47 | ±0.01 |
| Lactate (mmol/l) | 0.67 | 0.70 | 0.69 | ±0.02 |

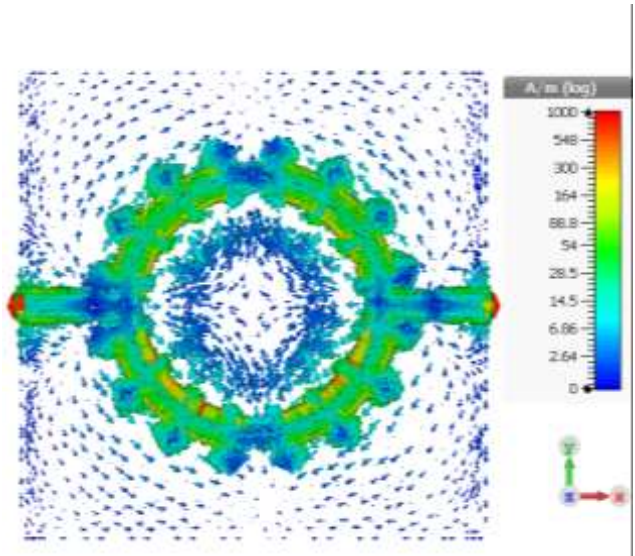

**FIGURE 14**: Surface current distribution across the coronavirus-shaped sensor at resonant frequency of 4.75 GHz.

Figure 15 shows that the electric field distributions over the proposed sensor at resonant frequency of 4.75 GHz are mostly located at the segment of discrete port one with a small concentration around port two. This is where the electric field distribution spanned towards port tow with less concentration around port one. It is known that the electromagnetic field can travel through the two ported conductive transmission lines, where the electric field and magnetic field components are perpendicular to each other and are perpendicular to the propagation direction. The transmission line is copper, which is a metallic layer and, hence, it is affected by the electric field component of the travelling wave.

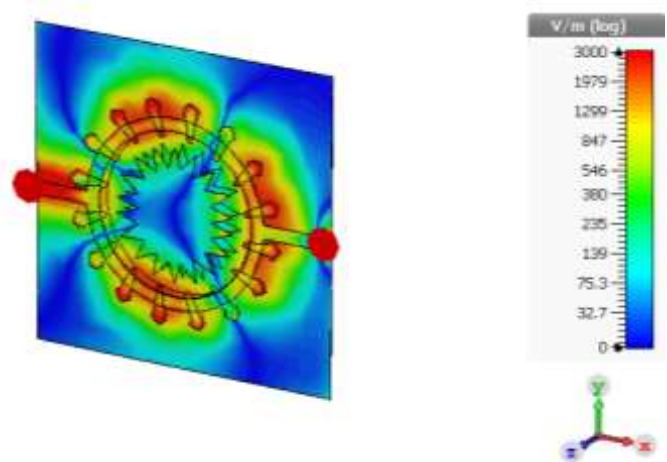

**FIGURE 15**: Electric field distribution across the coronavirus-shaped sensor at resonant frequency of 4.75 GHz.

## V. CONCLUSIONS

In this study and for the first time, a metamaterial-based sensor was successfully employed for rapid diagnosis of covid-19. The sensor was designed numerically and tested experimentally by evaluating variations in the reflection coefficient (S11) and transmission coefficient (S21) of the waves at resonant frequency. The measured electrical properties (dielectric constant and dielectric loss factor) of the normal blood are 64.9 and 50 respectively, Results of covid-19 relevant blood sample showed a pronounced shift in the main resonant frequency of about 740 MHz compared to that of the control blood sample. It was concluded that with the help of the proposed sensor a significant breakthrough can be achieved for rapid diagnosis of covid-19 within few seconds.

## Funding

This work was supported by the National Key Research andDevelopment Program of China (Grant no. 2017YFA0204600),the National Natural Science Foundation of China (Grantno. 51802352) and the Fundamental Research Funds for theCentral Universities of Central South University (Grant no.2018zzts355).

## Conflicts of interest

The authors declare no conflict of interest.

## Acknowledgments

The author would like to thanks for Central South Universityand Iskenderun Technical University for the technical sup-ports.

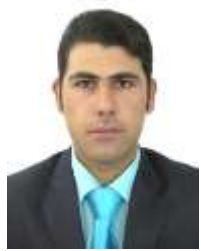

**YADGAR I.ABDULKARIM** was born in sulaimanyah - Kurdistan Region- Iraq , in 1984. He was awarded B.Sc. degree in Physics at Sulaimani University, Sulaimanyah, Kurdistan Region-Iraq in 2007. M.Sc.degree in theoretical Physics from V.N. Karazin Kharkiv National University Faculty of Physics - Academician I.M. Lifshyts Theoretical Physics Department, Kharkiv –Ukraine in 2012 .Ph.D in Condensed matter physics at Central South University-School of Physics and Electronics, Hunan Provience - Changsha city in 2020 . He has published more than 19 Articles. His research interests are metamaterials design, fabrication and their applications, microwave absorbing materials, Antenna, and wavguides.

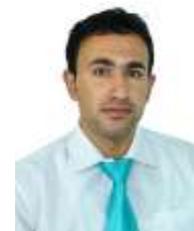

**HALGURD N.AWL** was born in sulaimanyah, Iraq , in 1988. He received the B.S. degree in Communication Engineering from the Sulaimani Polytechnic University in 2011 and M.S. degrees in Radio Freuency and Microwave engineering from the University of Birmingham in 2014 ,.

Since 2014, he has been an Lecturer with the Communication Engineering Department, Rngineering College ,Sulaimni Polytechnic University . He is the author of more than 10 articles. His research interests include Radio Frequency, Microwave, and THz components and device design such as antenna , filter , sensor , absorber .

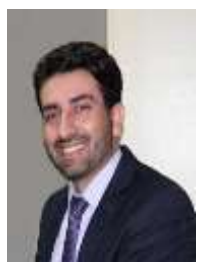

**FAHMI F. MUHAMMADSHARIF** received the B.S. degree in Physics from University of Sulaimani, Kurdistan Region, Iraq, in 2004, the M.S. degree in Physics from Mosul University, Iraq, in 2007 and the Ph.D. degree in Solar Energy from University of Malaya (UM), Malaysia, in 2012.

He was Postdoctorate Fellow and Visiting Researcher at Universiti Teknologi Malaysia (UTM) from 2016 to 2018. He is currently serving as Assistant Professor and Researcher with Koya University, Kurdistan Region, Iraq. Dr. Fahmi is the author of more than 40 ISI/WoS and conference papers. He is the member of International Solar Energy Society (ISES), International Association of Advanced Materials (IAAM) and Editorial of ARO-The Scientific Journal of Koya University. His awards and honors include PDF Fellowship of UTM, Malaysia, PhD Scholarship of Nanakali Foundation, Erbil, KRG-Iraq, ICFMD 2010 Gold Medal, and Research & Travel grants (PS319/2009B and PS343/2010B) from UM. His research interests are in the fields of Solar Energy, Organic Elecronics, Metamaterials and Machine Learning.

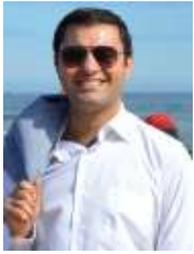
**KARZAN R. SIDIQ** was born in Chamchamal/Sulaimanyah, Kurdistan Region-Iraq in 1979. He was awarded Technical Diploma in community health at Sulaimani Technical Institute, Sulaimanyah, Kurdistan Region-Iraq in 2000; B.Sc. in biology at Sulaimani University, Sulaimanyah, Kurdistan Region-Iraq in 2004; M.Sc. in microbiology at Salahaddin University, Erbil, Kurdistan Region-Iraq in 2007; Ph.D. in molecular microbiology at Newcasle University, Newcastle Upon Tyne, United Kingdom in 2016. He has more than 10 years of experience in teaching/lecturing and research. He is currenly employed by Charmo University, Kurdistan Region-Iraq with the job title of LECTURER and RESEARCHER. He has seven published articles that are concerned with bacteria, antibiotics and viruses. He has research interest in bacteriology, virology antibiotic resistance, molecular biology and development of microbial diagnostic techniques.

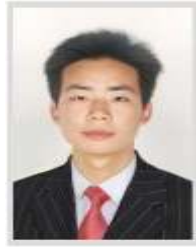
**HENG LUO** (1988-) has received Ph.D degree of materials physics and chemistry in Central South University at 2015. Research interests focus on the wireless power transmission technology, electromagnetic compatibility / interference (EMC/EMI), nanocarbon functionalized composites, etc. More than 30 peer-reviewed works are published on Applied Physics Letters, Scientific Reports, Journal of Applied Physics, Journal of Euronpean Ceramics Sociecty, CARBON, etc., and two national patents of invention have been authorized. His current research interests include wireless power transmission technology and antennas.

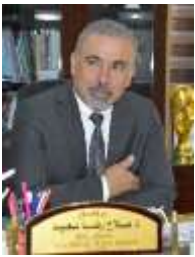
**SALAH R. SAEED** received the B.S. degree in Physics from Salahaddin University, Kurdistan Region, Iraq, in 1988, the M.S. degree in Physics from Salahaddin University, Kurdistan Region, Iraq, in 1992 and Ph.D. degree in Nanotechnology and Nanostructure from Jagiellonian University, Krakow, Poland, in 2008.

He was Postdoctorate Fellow from January to August 2009 at Jagillonian University, Krakow, Poland. He is currently serving as Full Professor and President of Charmo University, Kurdistan Region, Iraq. Prof. Salah is the author of more than 30 ISI/WoS and conference papers. His research interest includes Physics of Nanotechnology and Nanostructure.

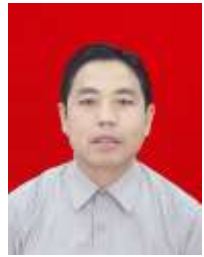
**LIANWEN DENG** received the B.S. degree and the Ph.D. degree in electrical engineering from Huazhong University of Science and Technology, Wuhan, China, in 1991 and 2004, respectively. From 2011 to 2012, he was a visiting scholar of the institute of microwave material and instrument, Colorado State University, USA. He is currently vice president of School of Physics and Electronics, Central South University, China. His current research interests include wireless power transmission technology, antennas, and integrated circuit design

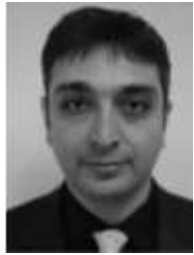
**MUHARREM KARAASLAN** received the PhD degree in Physics Department from University of Cukurova, Adana, Turkey, in 2009. He has authored more than 100 research articles and conference proceedings. His research interests are applications of metamaterials, analysis and synthesis of antennas, and waveguides.

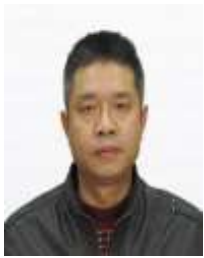
**SHENGXIANG HUANG** received his PhD degree from Central South University (CSU), China and joined CSU as a staff .Since then, he has been engaged in designing and fabrication of electronic materials , and developing microwave transmissions like antenna.

9